\documentclass{PoS}

\title{Status of SuperKEKB and Belle II}

\ShortTitle{Status of SuperKEKB and Belle II}

\author{\speaker{Thomas Kuhr} for the Belle II collaboration\\
        Institut f\"ur Experimentelle Kernphysik, KIT\\
        E-mail: \email{Thomas.Kuhr@kit.edu}}

\abstract{High precision flavor physics measurements 
are an essential complement to the direct searches for new physics at the LHC.
Such measurements can be performed at an upgraded KEKB accelerator.
The status of the SuperKEKB collider and Belle II detector is presented
in this article.}

\FullConference{Flavor Physics and CP Violation - FPCP 2010\\
		May 25-29, 2010\\
		Turin, Italy}

\begin{document}

\section{Introduction}
The B factory experiments, Belle at the KEKB collider at KEK \cite{Belle:2000cg} and
BaBar at the PEP II collider at SLAC \cite{Aubert:2001tu}, were built to measure the
large mixing-induced $CP$ violation in the $B^0$ system predicted by the theory of
Kobayashi and Maskawa \cite{Kobayashi:1973fv}.
The successful confirmation of the prediction led to the Nobel Prize for both theorists.

\begin{figure}
\centering
\includegraphics[width=\textwidth]{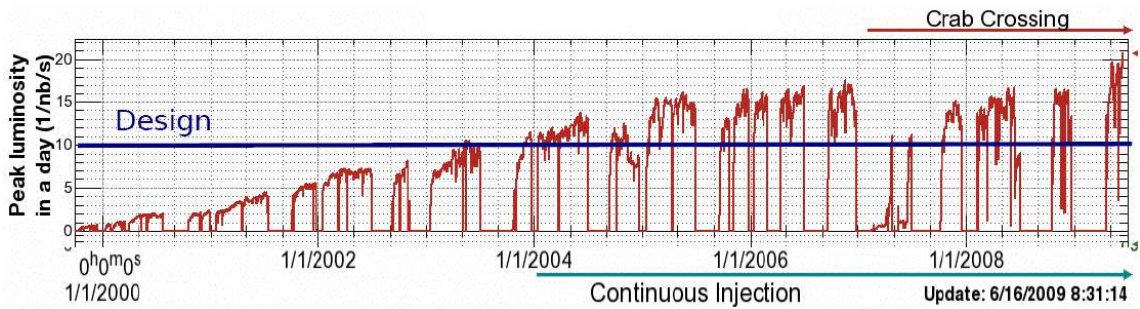}
\caption{Evolution of KEKB peak luminosity.}
\label{fig:lumi}
\end{figure}

In addition to the precise measurement of the CKM angle $\phi_1$, a broad physics program
is carried out at the B factories.
The output of physics results that exceeded initial expectations was supported by
the excellent performance of the accelerators.
With an instantaneous luminosity world record of $2.1 \times 10^{34}$~cm$^{-2}$s$^{-1}$,
more than twice the design luminosity (see Fig.~\ref{fig:lumi}),
the KEKB accelerator was able to deliver a total integrated luminosity of 1 ab$^{-1}$.
Technological innovations, like continuous injection, developed at SLAC, and crab
cavities, developed at KEK, contributed to this achievement.

Most B factory results are in good agreement with the expectations from the Standard Model (SM)
and confirm the CKM structure of quark mixing and $CP$ violation, 
but some measurements show tensions with the SM prediction.
Much larger datasets are needed
to investigate whether these are first hints for effects of New Physics (NP) models.
The task of the super flavor factories, like SuperKEKB, is
to acquire such high-statistics datasets for high-precision measurements
that allow to search for significant deviations from the SM
which are expected to exist.

\section{Physics Case}
Although the Standard Model is very successful in describing the various
current experimental results, it has theoretical issues at energies studied at the LHC.
Moreover it fails, for example, to explain the asymmetry between matter and anti-matter observed
in the universe which would require an additional source of $CP$ violation.
The SM is thus believed to be a low-energy approximation of a more fundamental theory.
A complementary approach to direct searches for new particles at the LHC is the search
for deviations from the SM predictions due to virtual contributions of new particles
in precision flavor physics measurements.
This is the main physics objective of the Belle II experiment at the SuperKEKB accelerator.
If the direct searches for NP at the LHC are successful, the precision flavor physics measurements will provide 
essential information to identify the kind of NP.
This feature is sometimes called ``DNA test'' of NP 
\cite{Altmannshofer:2009ne}.

\begin{figure}
\centering
\includegraphics[width=0.45\textwidth]{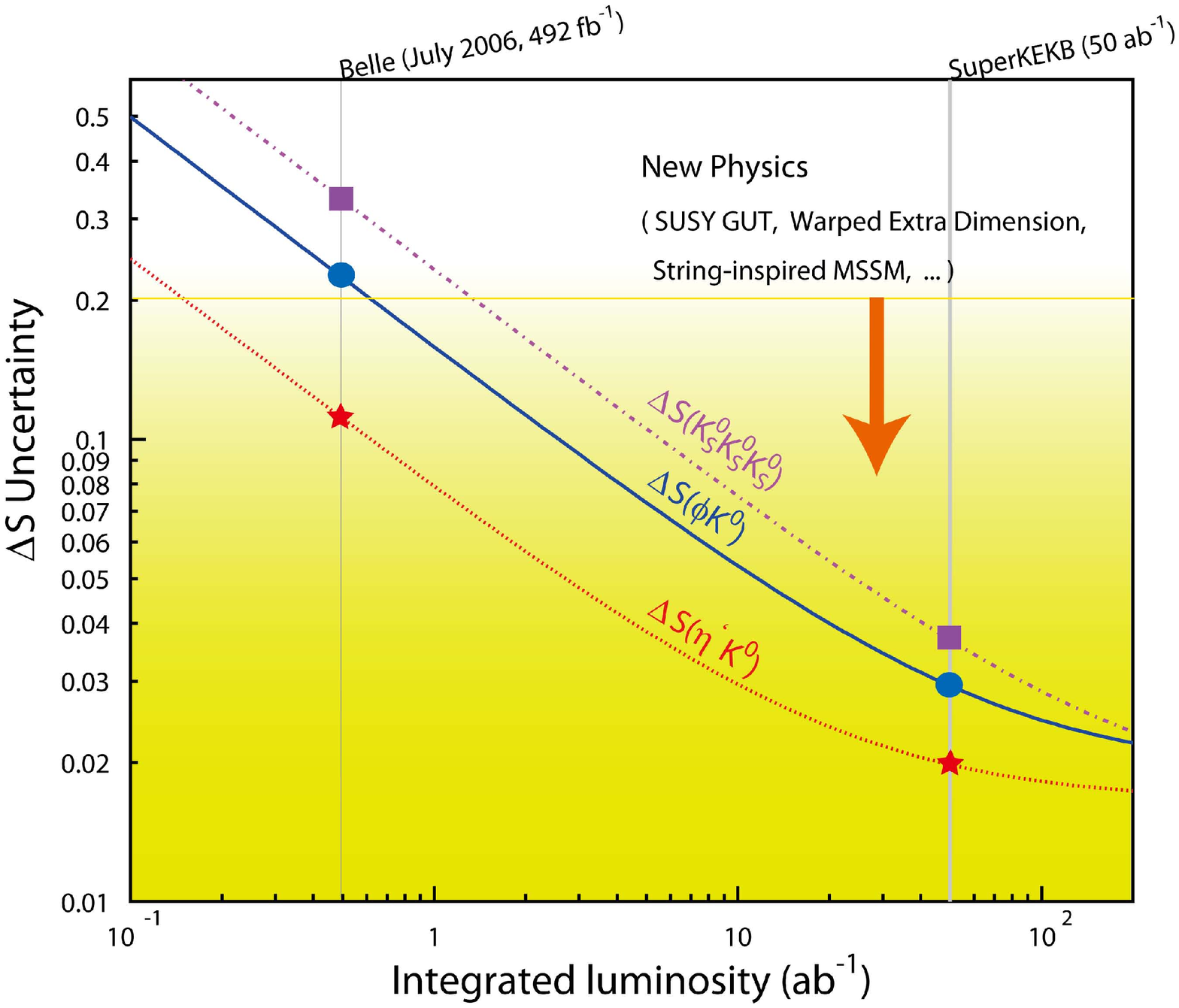}
\hspace*{1cm}
\includegraphics[width=0.45\textwidth]{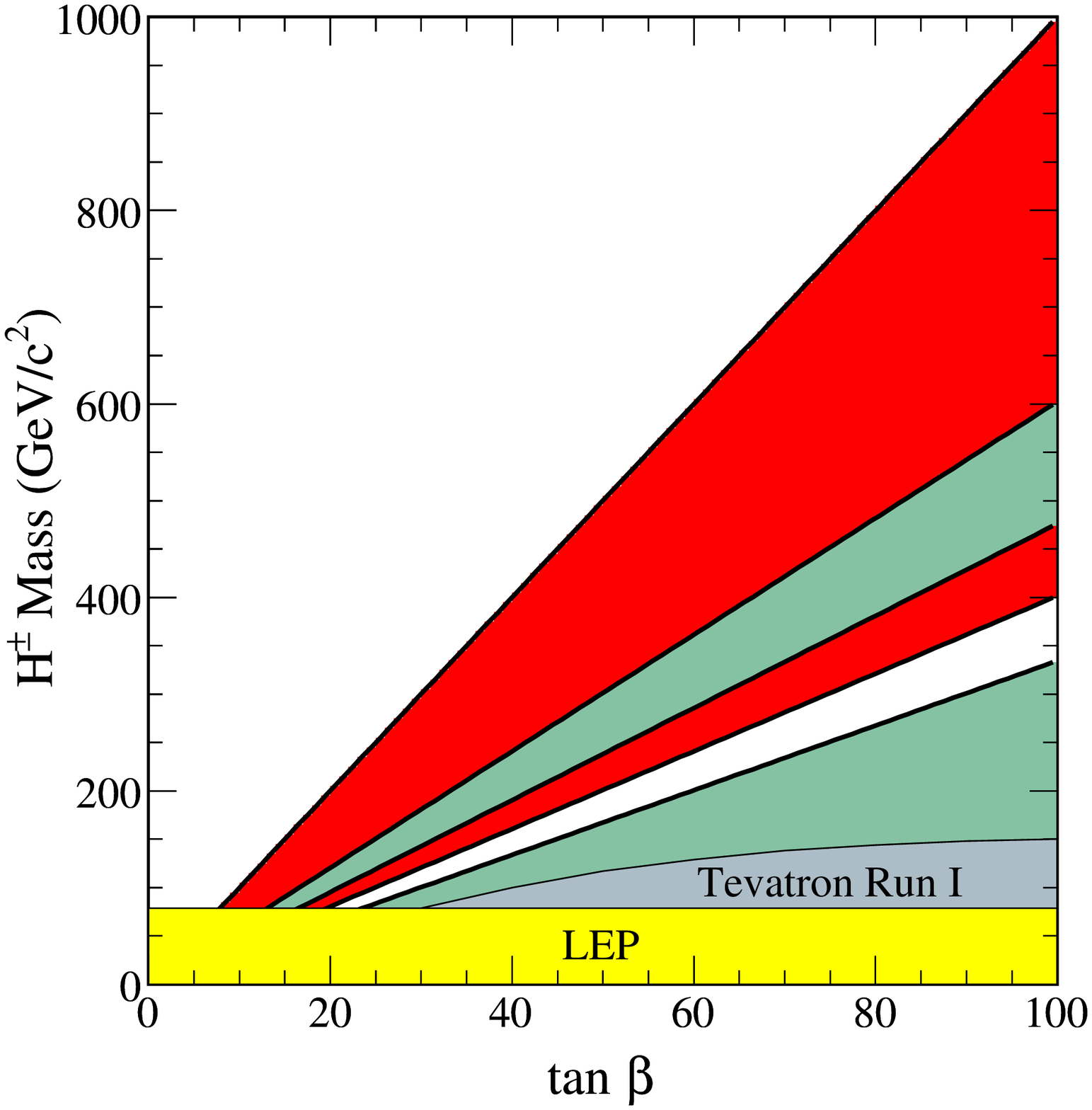}
\caption{Left: Projection of the sensitivity to NP contributions to mixing induced
$CP$ violation in $b \rightarrow q\bar{q}s$ decays.
Right: Excluded values of charged Higgs mass as a function of $\tan\beta$ (yellow, blue, and green) 
and 5$\sigma$ discovery region for 5 ab$^{-1}$ for $B \rightarrow \tau\nu$ (red).}
\label{fig:b_qqs}
\end{figure}

In the past a discrepancy between the mixing induced $CP$ violation in $b \rightarrow c\bar{c}s$ and
$b \rightarrow q\bar{q}s$ decays was observed.
Although this discrepancy is resolved by the current measurements, their uncertainties are
still large enough to hide sizable NP contributions.
Since the precision of most measurements is currently limited by their statistical uncertainties,
a dataset of 50 ab$^{-1}$, as aimed for at SuperKEKB, would allow to either restrict the
parameter space of NP models considerably or to find NP effects, until becoming limited by
theoretical uncertainties as shown in Fig.~\ref{fig:b_qqs}.

The unitarity of the CKM matrix in the SM is often depicted by the unitarity triangle.
A precise measurement of the angles of the triangle is therefore essential for testing unitarity.
The anticipated precision achievable with 50 ab$^{-1}$ is 0.012 for $\sin\phi_1$, 
$\sim1^\circ$ for $\phi_2$, and $1.5^\circ$ for $\phi_3$.

A possible hint of NP might have been seen in the forward-backward (FB) asymmetry in 
$B \rightarrow K^*\ell^+\ell^-$ decays.
BaBar, Belle, and CDF measurements show a trend towards values above the SM prediction.
To establish evidence for NP or confirm the SM, the invariant di-lepton mass
where the FB asymmetry vanishes, has to be precisely known.
The zero crossing point, which is less model dependent than the overall shape, can be
determined with 5\% accuracy with 50 ab$^{-1}$.

\begin{figure}
\centering
\includegraphics[width=0.45\textwidth]{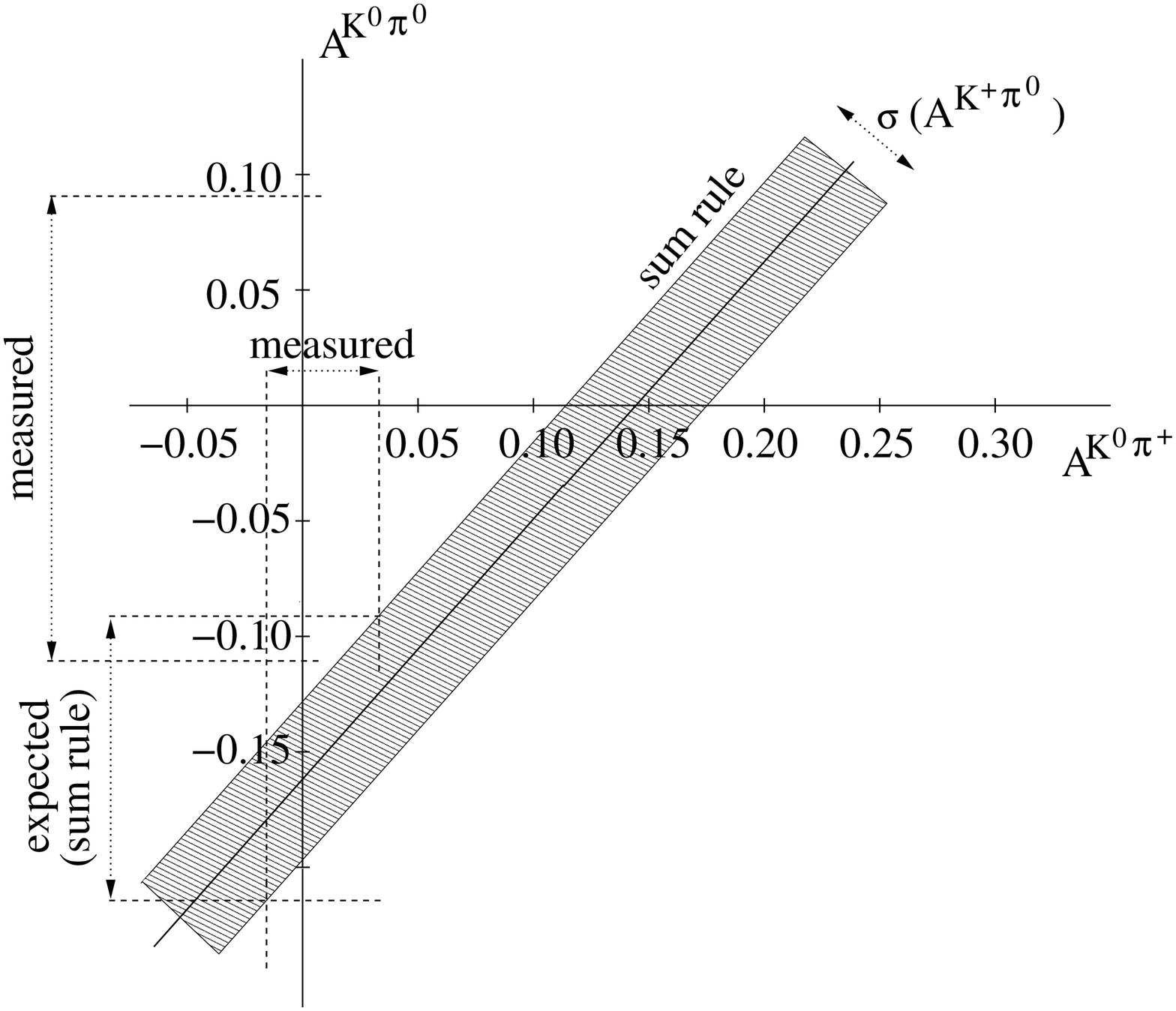}
\hspace*{1cm}
\includegraphics[width=0.45\textwidth]{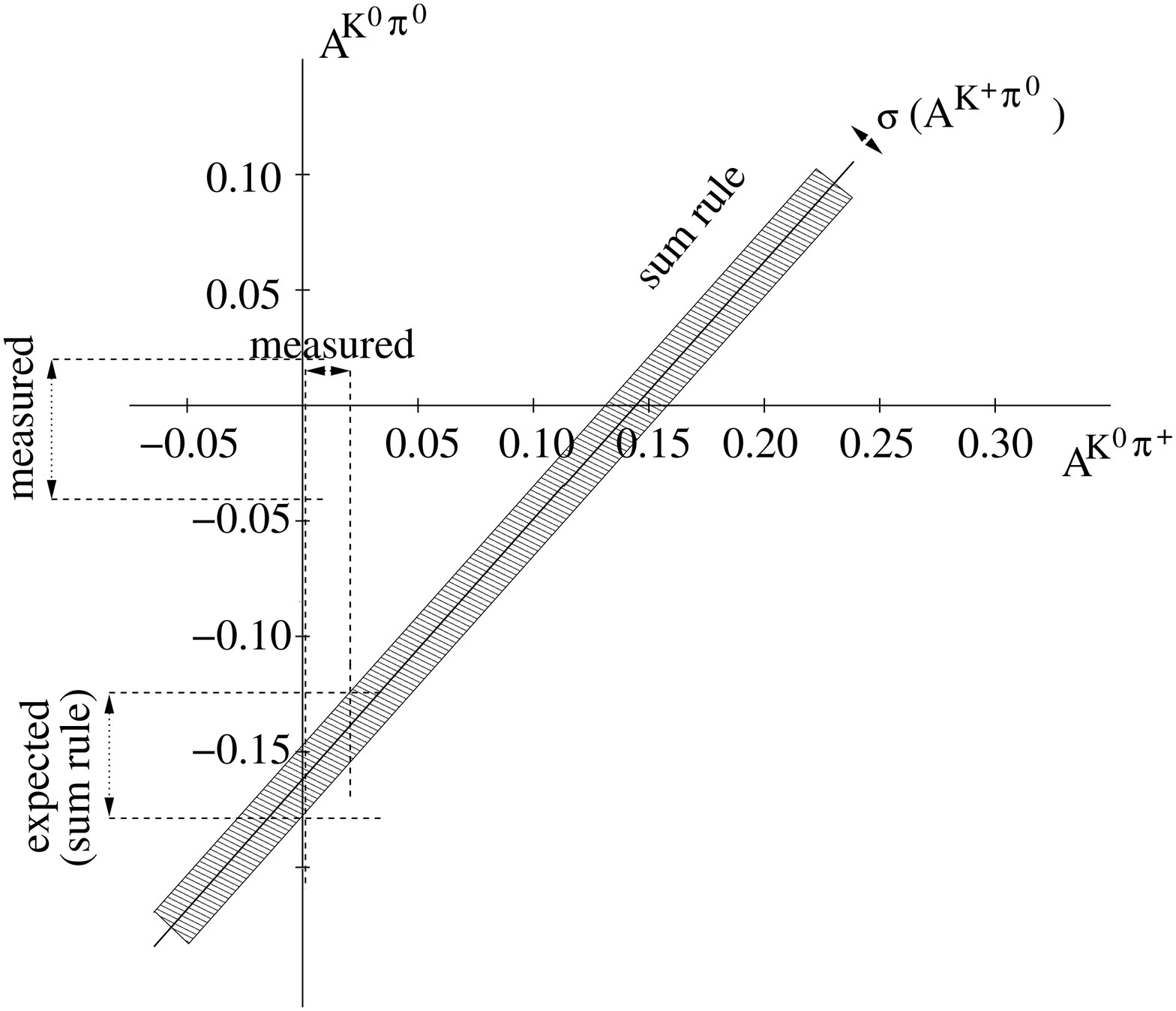}
\caption{Illustration of the sum rule proposed in Ref.~\cite{Gronau:2005kz} for the
current experimental values (left) and the projection for SuperKEKB assuming the same
central values (right).}
\label{fig:BKpi}
\end{figure}

Another unexpected result is the difference between direct $CP$ violation in 
$B^0 \rightarrow K^+\pi^-$ and $B^+ \rightarrow K^+\pi^0$ decays.
To check whether sub-leading SM contributions can account for this discrepancy,
the validity of a sum rule, involving the $CP$ asymmetry in $B^0 \rightarrow K^0\pi^0$ 
and $B^+ \rightarrow K^0\pi^+$ decays, as suggested in \cite{Gronau:2005kz} has
to be tested.
Figure \ref{fig:BKpi} illustrates the expected improvement in the precision of this test.

A further tension has recently grown in $B^+ \rightarrow \tau^+\nu_\tau$ decays.
The branching ratios measured by BaBar and Belle are more than 2 standard deviations
higher than the SM expectation.
This could be caused by contributions from a charged Higgs.
Figure~\ref{fig:b_qqs} shows the discovery potential for a dataset of 5 ab$^{-1}$.
Significant improvements in sensitivity are also expected in searches for
lepton flavor violation and $CP$ violation in $D^0$ decays.
A detailed discussion of the physics program at SuperKEKB can be found in
Ref.~\cite{Aushev:2010bq}.

Last but not least, unexpected effects that were not yet considered may show up
when increasing the statistics by about two orders of magnitude.
The discovery of the $X(3872)$ that triggered great interest in exotic hadron
spectroscopy on experimental and theoretical side can be regarded as an example
for such an unexpected effect at the predecessor experiment Belle.

\section{SuperKEKB Accelerator}
\begin{figure}
\centering
\includegraphics[width=0.7\textwidth]{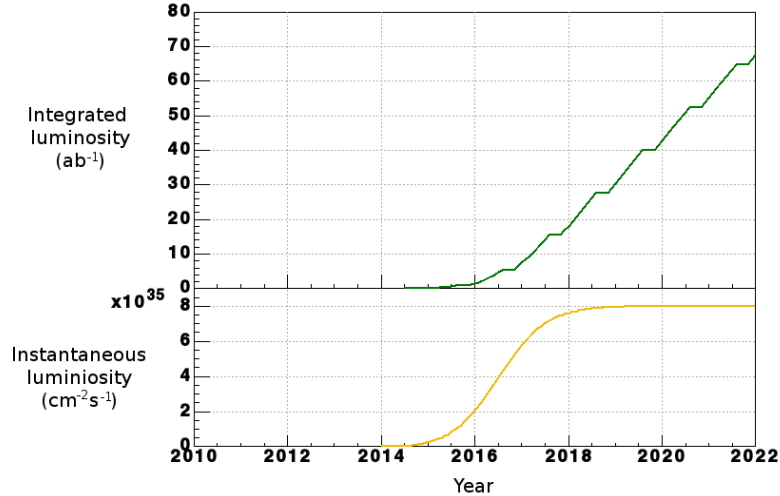}
\caption{Schedule of planned SuperKEKB performance.}
\label{fig:projection}
\end{figure}

The large statistics required to reach the physics goals can only be achieved by a new
generation of $e^+e^-$ colliders.
Aiming for 50 ab$^{-1}$ in the year 2020, a design luminosity of 
$8 \times 10^{35}$~cm$^{-2}$s$^{-1}$ is required for SuperKEKB.
The projected instantaneous and integrated luminosity is shown in Fig.~\ref{fig:projection}.

The luminosity $\mathcal{L}$ is given by
\begin{equation}
\mathcal{L} = \frac{\gamma_\pm}{2er_e} \left(1 + \frac{\sigma_y^*}{\sigma_x^*} \right)
\frac{I_\pm \xi_{y,\pm}}{\beta^*_{y,\pm}} \frac{R_L}{R_{\xi_y}}
\label{eqn:lumi}
\end{equation}
where $\gamma$ is the Lorentz factor, $\sigma_y^*/\sigma_x^*$ the beam size aspect ratio, 
$I$ the beam current, $\beta^*_y$ the vertical beta function at the interaction point,
$\xi_y$ the beam-beam parameter, and $R_L/R_{\xi_y}$ a geometrical factor.
The subscript $\pm$ refers to the product of the corresponding quantities for the 
low energy positron (LER) and high energy electron (HER) beams.

To reach a forty times higher luminosity than at KEKB, the contributions of several improvements
have to be added.
The main increase in luminosity comes from a significantly smaller beam size at the
interaction point (nano beam scheme).
The beta functions are reduced in $y$ direction from 5.9 mm to 0.27/0.42 mm for HER/LER, and in $x$
direction from 120 cm to 3.2/2.5 cm.
For this a new interaction region is designed with new focusing quadrupole magnets.

Since the beam-beam parameter is proportional to $\sqrt{\beta^*/\epsilon}$, the emittance $\epsilon$
has to be reduced to keep the beam-beam parameter at the same level as at KEKB.
A reduction of the emittance from 18/24 nm to 3.2/1.7 nm is obtained by installing a new
electron source and a new damping ring, in addition to a redesign of the HER arcs.
The last contribution to the luminosity gain comes from higher beam currents.
They are increased from 1.6/1.2 A to 3.6/2.6 A.

The higher luminosity also leads to higher background levels.
At SuperKEKB Touschek scattering becomes the dominant background source.
Furthermore the design for the luminosity of $8 \times 10^{35}$~cm$^{-2}$s$^{-1}$
requires to reduce the beam energy asymmetry from 3.6/8 GeV to 4/7 GeV
and to enlarge the crossing angle from 22 mrad to 83 mrad.

\section{Belle II Detector}
\begin{figure}
\centering
\includegraphics[width=\textwidth]{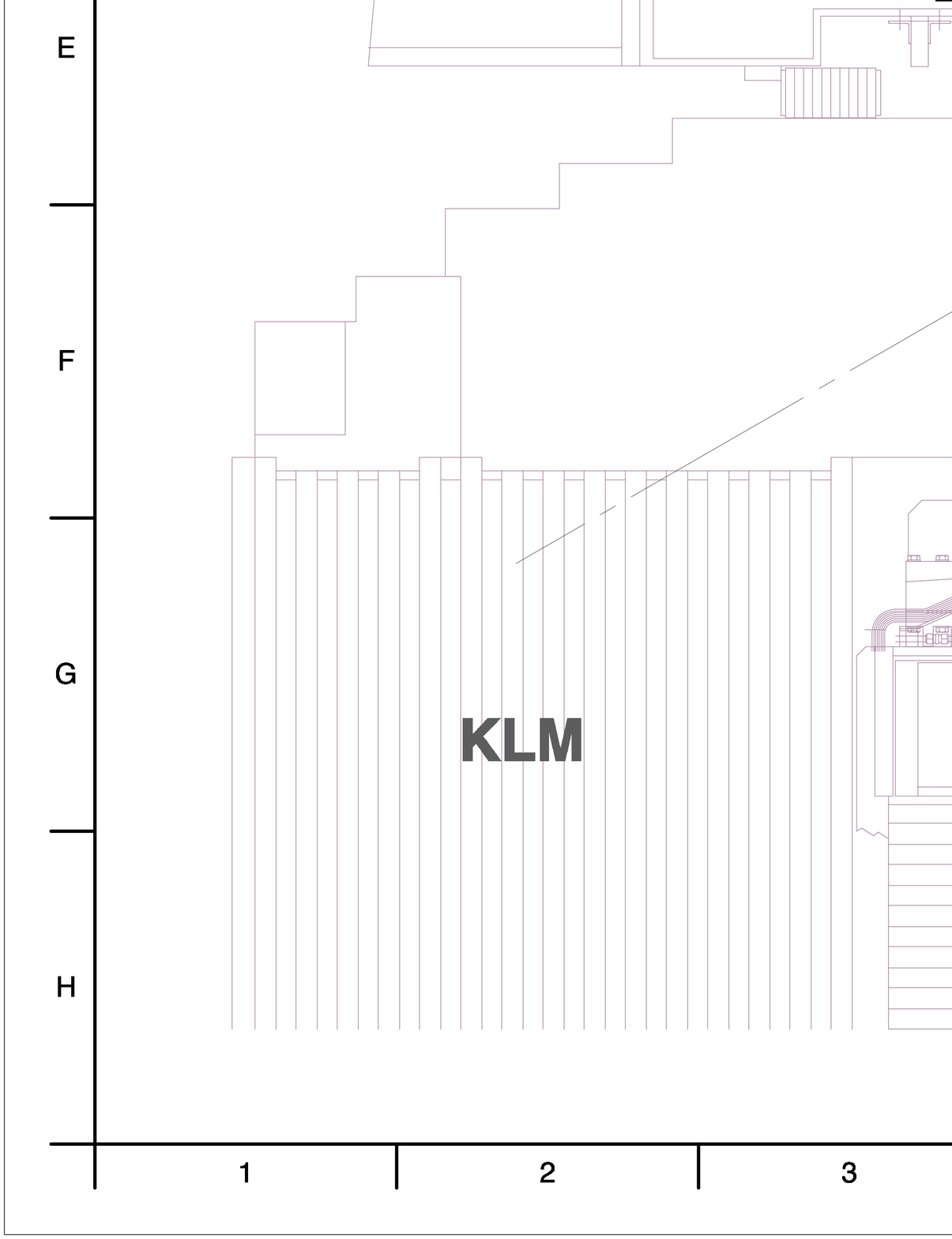}
\caption{The Belle II detector (top half) compared with the Belle detector (bottom half).}
\label{fig:detector}
\end{figure}

Because of the increased background level, the Belle II detector has to deal with
higher occupancy and radiation damage.
In addition the increased event rate puts high demands on trigger, data acquisition, and computing.
To cope with the conditions at the SuperKEKB collider, the components of the Belle
detector are either upgraded or replaced by new ones.
Figure~\ref{fig:detector} shows a comparison of the Belle and Belle II detectors.

The innermost part of the tracking system consists of two layers of silicon pixel
sensors (PXD) based on the DEPFET technology.
It is surrounded by four layers of double sided silicon strip detectors (SVD).
With the excellent spatial resolution of the PXD an impact parameter resolution in beam direction
of $\sim20$ $\mu$m can be achieved leading to an improved determination of the vertex position.
The larger outer radius of the SVD compared to Belle gives an increase in efficiency
of about 30\% for the reconstruction of $K_S \rightarrow \pi^+\pi^-$ decays inside the SVD.
A precise measurement of the momentum of charged tracks is provided by the central
drift chamber (CDC).
Improvements in the momentum resolution compared to the Belle CDC are achieved by a
larger outer radius and a smaller cell size.

For the identification of charged hadrons, the time-of-flight detector at Belle is replaced
by a time-of-propagation counter (TOP).
The usage of timing information of internally reflected Cherekov light allows for a compact
design of this particle identification device in the barrel part.
The forward region is instrumented with new RICH detectors (ARICH) using aerogel layers with different
refractive index to generate Cherekov rings with the same radius for each layer.
A kaon identification efficiency of $>99$\% (96\%) at a pion mis-identification rate
of $<0.5$\% (1\%) is expected for $B \rightarrow \rho\gamma$ events reconstructed in the
TOP counter (4 GeV particles reconstructed in the ARICH).

The crystals of the Belle electromagnetic calorimeter (ECL) will be reused for Belle II.
A replacement with faster and more radiation tolerant crystals in the endcap region is
considered as upgrade option.
To improve the signal to background separation under the higher background conditions at SuperKEKB,
the electronics will be upgraded to enable a wave form sampling.
Muons and $K_L$ mesons are identified by resistive plate chambers in the outer part
of the Belle detector (KLM).
For Belle II the endcap regions will be upgraded with scintillator strips to cope with
the higher background rates.

\begin{figure}
\centering
\includegraphics[width=0.8\textwidth]{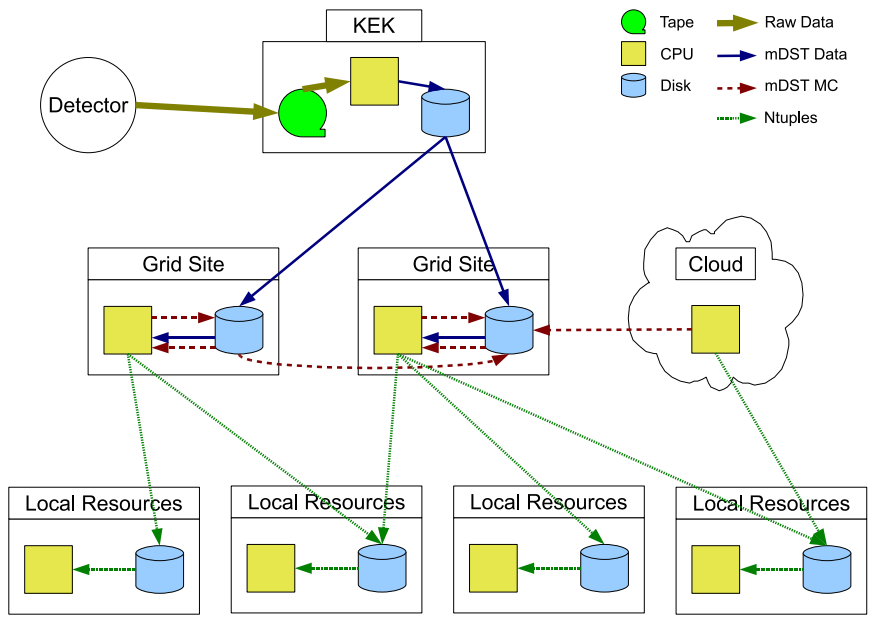}
\caption{The Belle II computing model.}
\label{fig:computing}
\end{figure}

The almost two orders of magnitude higher rate of interesting physics events requires
to upgrade the data acquisition system and the offline computing system.
Both will use a common software framework with ROOT as persistency layer.
In contrast to the KEK centric computing model of Belle, the MC production and physics
analysis at Belle II will be done in a distributed way exploiting grid and cloud technologies
as shown in Fig.~\ref{fig:computing}.

\section{Project Status}
In March 2008 a first proto collaboration meeting was held.
In December of the same year the Belle II collaboration was officially founded.
Currently it has about 350 members from over 50 institutes and it is steadily growing.
With members from 13 different countries in Asia, Europe, America, and Australia
it is a truly international project.

After a preliminary approval of the SuperKEKB upgrade by the
Japanese Government in January 2010, the ministry that supervises KEK 
assigned in June 100 oku-yen (approx \$110M) to the project 
from the "Very Advanced Research Support Program".
This covers about one third of the total estimated cost.

The Belle experiment has stopped data taking and the KEKB accelerator was switched
off on June 30th 2010 to start the upgrade.
It is planned to take first data with Belle II at SuperKEKB in 2014.
The design luminosity should be reached in 2018 so that a total data sample
of 50 ab$^{-1}$ can be accumulated until 2020.

\section{Summary}
Deviations from the SM, which is known to have shortcomings, may still hide in
flavor physics observables.
To explore this territory, high-precision measurements are needed.
A next generation flavor factory, like SuperKEKB, can provide the required
high-statistics data samples.
It is designed for a luminosity of $8 \times 10^{35}$~cm$^{-2}$s$^{-1}$,
aiming for an integrated luminosity of 50 ab$^{-1}$ in 2020.
The Belle detector components are either upgraded or replaced by new ones
to cope with the more challenging beam conditions and to improve the
detector performance.
The design of the accelerator and detector are described in detail in the
Belle II Technical Design Report that will be published soon.
Overall the project is well on track to start data taking in 2014.

\end{document}